\documentclass[9pt,twocolumn,twoside,lineno]{pnas-new}
\usepackage{amsmath}
\usepackage{amssymb}
\usepackage{amsthm} 
\usepackage{svg}
\definecolor{nicegreen}{rgb}{0, 0.593, 0}
\usepackage{xcolor}
\usepackage{graphicx}
\usepackage{xcolor}
\usepackage{comment}
\usepackage{tcolorbox}
\usepackage{fancybox}
\usepackage{svg}

\newtheorem{remark}{Remark}

\templatetype{pnasresearcharticle} 

\title{Dual Pricing to Prioritize Renewable Energy and Consumer Preferences in Electricity Markets}

\author[a]{Emilie Jong}
\author[b]{Samuel Chevalier}
\author[a]{Spyros Chatzivasileiadis} 
\author[c,d,1]{Shie Mannor}

\affil[a]{Technical University of Denmark}
\affil[b]{University of Vermont}
\affil[c]{Technion}
\affil[d]{Jether Energy Research}

\leadauthor{Jong} 

\significancestatement{This paper introduces a framework which allows consumers to place a bid for black and green energy in the day-ahead market. Due to transmission lines, renewable energy often gets curtailed, preventing its full dispatch despite the high demand. By considering network topology and transmission constraints, dispatching energy from black generators can alleviate these bottlenecks, allowing more green energy to flow. Letting consumers bid on green energy enables payments to black generators for congestion relief, and thus promotes additional green energy dispatch. Using this framework, we derive prices for green and black energy.}

\authorcontributions{S. Chevalier, S. Chatzivasileiadis and S.M. designed research; S. Chevalier and E.J. performed analysis; S. Chevalier and E.J. wrote the paper; S. Chatzivasileiadis and S.M manuscript editing}
\authordeclaration{The authors declare no competing interest.}
\correspondingauthor{\textsuperscript{1}To whom correspondence should be addressed. E-mail: shie@ee.technion.ac.il}

\keywords{Electricity markets $|$ Locational Marginal Pricing $|$ Curtailment} 

\begin{abstract}
Electricity markets currently fail to incorporate preferences of buyers, treating polluting and renewable energy sources as having equal social benefit under a system of uniform clearing prices. Meanwhile, renewable energy is prone to curtailment due to transmission constraints, forcing grid operators to reduce or shut down renewable energy production despite its availability and need. This paper proposes a ``dual pricing mechanism" which allows buyers to bid both their willingness to pay for electricity, and additionally, their preference for green energy.
Designed for use in deregulated electricity markets, this mechanism prioritizes the dispatch of more renewable energy sources according to consumer preferences. Traditional uniform clearing prices, which treat all energy sources equally, do not reflect the growing share of green energy in the power grid and the environmental values of consumers. By allowing load-serving entities to bid their willingness to pay for renewable energy directly into the clearing market, our proposed framework generates distinct pricing signals for green and ``black" electricity. 
\end{abstract}

\dates{This manuscript was compiled on \today}
\doi{\url{www.pnas.org/cgi/doi/10.1073/pnas.XXXXXXXXXX}}

\begin{document} 

\maketitle
\thispagestyle{firststyle}
\ifthenelse{\boolean{shortarticle}}{\ifthenelse{\boolean{singlecolumn}}{\abscontentformatted}{\abscontent}}{}

\dropcap{E}lectricity markets give consumers limited influence over selecting the origin of the electricity source powering their homes. In many electricity markets across the world, the market is cleared at a single uniform market clearing price under the ultimate objective of maximizing social welfare. In the absence of transmission constraints, the price of electricity is thus {\em identical} for all sources, regardless of their origins. This market structure is in place because electricity is considered a commodity, i.e., a good that is two-way interchangeable. Once electricity is integrated into the grid, its source becomes indistinguishable to consumers. This pricing mechanism 
assumes that all units of electricity are societally valued at the same price, and all electricity units are treated equally without consideration of production methods. In reality, clearing prices might differ at certain locations due to constraints in the network, leading to curtailment of renewable energy sources. When considering transmission constraints, clearing prices can be determined by zonal or nodal pricing structures~\cite{eicke2022}. In a nodal pricing structure, as in the US~\cite{Joskow:2019}, clearing prices vary from node to node in a network; in the case of zonal pricing, as in the EU~\cite{bolton2021making}, prices vary among geographical bidding zones. In this structure, the capacity of intra-zonal transmission lines is neglected, establishing a singular zonal clearing price~\cite{eicke2022}.

In this paper, we propose a market framework where 
consumers differentiate between green and black electricity sources through bidding their preference in a dual bid. This approach generates distinct pricing signals for renewable and non-renewable energy, prioritizing renewable sources in the dispatch process in accordance with market preferences. Assuming customers are willing to pay more for green power, this shift will not only promote additional green energy dispatch and deployment, but it will also help align consumer preference with support for a sustainable energy future.

\subsection*{Deregulated electricity markets}

Over previous decades, the evolution of electricity markets has been driven by liberalization, efforts to reduce carbon emissions, and rapid technological advancements. Prior to the 1980's, electrical power systems worldwide did not resemble economically efficient free markets~\cite{perez2014regulation,Cramton:2017}. Instead, ``vertically integrated'' monoliths centrally controlled the production, transmission, and selling of electrical power. In the face of rising energy costs, countries such as Chile, the UK, and Norway took the unprecedented step of \textit{deregulating} their electricity markets~\cite{Pinson:2023,SERRA2022100798,bolton2021making}. The purpose of deregulation (also known as liberalization, or restructuring) was to separate naturally monopolistic processes, such as the management of the transmission grid, from ones which would benefit from competition, such as retail and generation (i.e., the buying and selling of power)~\cite{Pinson:2023}. Today, much of North America, most of Europe, and many other power systems throughout the world are deregulated, meaning efficient, Walrasian wholesale markets~\cite{Moshe:2014} control the buying and selling of electrical energy. At the core of economically efficient free markets lies the principle of social surplus maximization. This principle ensures that the market operates at the equilibrium, where marginal social costs are equal to marginal social benefits ~\cite{smith1776inquiry, besanko:2020}. 

While deregulation has taken many forms, marginal pricing provides its crucial economic heartbeat~\cite{schweppe1988spot,energinet:2022}. Marginal pricing signals, often known as locational marginal prices (LMPs), represent the cost of supplying an additional unit of power at a specific location in the grid at a particular point in time. LMPs form a key concept in electricity markets and determine payouts for market participants, offer critical investment indicators, and produce a spatially and temporally detailed pricing map based on the electricity grid at various levels of precision. While the market is cleared at a single uniform clearing price when transmission constraints are absent, in reality, the flow of electricity is actually restricted to the capacity of the transmission lines. When these constraints are considered, driven by congestion, LMPs can be staggeringly different across various regions of a power system whose market is cleared with a uniform pricing philosophy (i.e., the counterfactual prices \textit{would have been} uniform in the absence of transmission constraints). LMPs can be calculated at the hyper-local nodal level or at the aggregated zonal/country level, and they capture both the real-time cost of electrical energy and the bottle-necking effect of transmission line constraints.

Despite the benefits of deregulation, storm clouds are brewing around its sacred core tenant of marginal pricing. As recently illustrated by the Danish system operator, Energinet, power systems are rapidly changing~\cite{energinet:2022}: they are moving from systems dominated by high marginal costs (coal, oil and gas) to one dominated by low, or even zero, marginal costs (solar and wind). In 2022, the war in Ukraine caused record-high electricity prices \cite{ieaelectricityreport}. Despite the fall in wholesale gas and electricity prices since the end of 2022, electricity prices rose in 2023 to around two times their historical levels \cite{eureport}. Electricity prices are expected to fall again in 2024; however, the gas market is expected to remain tight until 2025 due to the cut in Russian gas supply \cite{ieaelectricityreport, eureport}.  Politicians who are supportive of renewable energy roll-out are becoming frustrated with this paradigm, with leaders from the UK, France, and the European Commission all recently expressing the same sentiment: top marginal gas-generator prices should not be paid to renewable producers, whose marginal costs are zero, nor should these prices be passed along to consumers~\cite{Griffin:2022,Joseph:2022,Hirth:2022}.



In order to address these concerns and the enduring war in Ukraine, the EU recently passed a series of market reforms~\cite{EU_reform}. These reforms focus on incentivizing Purchase Power Agreements (PPAs) and two-way Contracts for Difference (CfDs)~\cite{EU_reform_explained}. In two-way CfDs, electricity producers are guaranteed a minimum selling price, but they must also return any profits above a given price cap to consumers. This helps consumers and producers hedge against price volatility, but it also potentially disincentivizes renewable producers from investing in the technologies that will allow them to produce when power is most needed~\cite{EU_reform_crit}, based on traditional market signals. Thus, while the EU market reforms may forestall price volatility in the short-term when gas prices are high, long-term investment signals could become muddled, and consequently, the European power grid might come to lack the investments it most desperately needs to fully embrace the green transition. 

The US, which has shown an enduring commitment to market-based remedies for power system challenges~\cite{hogan_reaction:2021}, has encouraged renewable energy integration primarily through state-run renewable portfolio standards (RPSs)~\cite{RPS:2021}. In an RPS, each unit of green energy produced has an associated renewable energy credit (REC) -- when a wholesale consumer purchases and retires this REC, they can claim their associated energy usage as ``green''. State-level RPS programs were embraced largely due to the national-level failure of Obama-era cap and trade legislation~\cite{cap_n_trade_fail}. Much research has sought to quantify the successes of these programs; a comprehensive University of Chicago study found that RPSs increase renewable production by a marginal 4.2\%, but they increase consumer costs by 17\%. The study concluded that ``RPS programs do reduce emissions, but at a high cost''~\cite{greenstone2019renewable}. More fundamentally, critics argue that RPS programs allow for ``greenwashing''~\cite{vt_digger_greenwashing}, giving some consumers the appearance of sustainability, but providing little consequence for the actual power grid (i.e., REC purchasing does not change operator dispatch decisions, and it does not consider whether the power was even deliverable to its destination, based on physical grid constraints).

\subsection*{Renewable energy integration policy deficiencies}
Despite the US and Europe's commitments to renewable energy integration, their strategies have focused on centralized policies which either mandate renewable energy usage (RPS, renewable energy directives, etc.), subsidize its existence (production and investment tax credits, CfDs, etc.), or penalize competitors' emissions (emissions trading system, cap and trade, etc.). However, neither region has embraced a market-based strategy which gives wholesale consumers (i.e., load serving entitites) a \textit{direct choice} in what they pay for (i.e., something to bid for). Such a strategy could be directly embedded inside day-ahead and real-time/intraday power system markets, where producers (generators) submit offers, consumers (loads) submit bids, and a centralized agent clears the market\footnote{This process is how the majority of deregulated power system markets operate. See, e.g., the Independent System Operator of New England~\cite{ISONE}, the Electric Reliability Council of Texas~\cite{ERCOT}, or NordPool~\cite{Nordpool}, which clears much of the European market.}. Most likely, such a straightforward strategy has not been embraced due to two salient, but flawed, assumptions:
\begin{enumerate}
    \item Since renewable resources have the lowest marginal costs, they are always dispatched ``first'' when the merit order curve is cleared~\cite{Hirth:2022}. Thus, operators may assume that consumer preference for renewables will have little-to-no impact on dispatch decisions and, consequently, market operations.
    \item As noted, the prevailing economic wisdom for consumers who bid into an electricity market is that ``an electron is an electron''. Thus, as long as loads are served, renewables are subsidized, and externalities are penalized, operators may assume that consumer preference will not be a relevant factor within a market clearing context.
\end{enumerate}
These assumptions hold in practicality when a power system's transmission network acts like a ``copper plate''~\cite{Coffrin:CP}, i.e., it has infinite network capacity and never experiences congestion constraints or losses. In such networks, zero-marginal-cost renewables are always dispatched at their full production offer, so preference of renewables has no impact on dispatch decision or prices. However, modern power grids are not copper plates; instead, network congestion is a major driver of highly volatile LMPs and renewable curtailment. 

\subsection*{Congested transmission networks} The amount of renewable energy curtailment in modern power systems is staggering~\cite{C-E:2022}. In 2022, California curtailed 2.4TWh\footnote{For reference, this is enough energy to power the state of Vermont for almost half a year~\cite{vt:2021}.} of wind and solar~\cite{CA_crutailment}.  In March of 2022, Texas was curtailing an average of 2.4 GWh of solar and wind each hour~\cite{ercot_curtailment}. And in 2018, Hydro-Québec spilled 10.4 TWh worth of water due to a lack of transmission capacity~\cite{HQ_curtailment}. Some portion of curtailed renewable generation can be unavoidable (e.g., in ERCOT, if renewables instantaneously produce more power than the loads can consume, or if generic transmission constraints (GTCs) block production due to stability constraints), but another portion is curtailed in pursuit of market objectives. For example, in September of 2021, when energy prices dipped negative, 3GW of wind and solar in Australia curtailed their own production~\cite{aus_curtail}. Similarly, in April of 2023, balancing prices across 8 Nordic bidding zones briefly dropped to -2200 Euros/MWh, exposing renewables to extreme price volatility~\cite{nordic_curtail}.


While congestion cannot be identified as the sole culprit in these cases, negative pricing signals and renewable curtailment are generally functions of \textit{market design}, where the ultimate goal is surplus maximization. Unnecessary renewable curtailment might suggest that market regulations use existing infrastructure inefficiently (i.e., from a societal perspective).
In the day-ahead market clearing context, the DC power flow~\cite{spyros:notes} based function which maximizes market surplus (or social welfare) typically takes the following form~\cite{zms:Nordpool,Hobbs:zms}:
\begin{align*}
\max_{p_{{\rm load}},p_{{\rm gen}}}\quad & c_{{\rm load}}^{T}p_{{\rm load}}-c_{{\rm gen}}^{T}p_{{\rm gen}} \numberthis\label{eq: zms}\\
{\rm s.t.}\quad\;\;\; & {\small \text{(i) line flow constraints}}\\[-2.5pt]
 & {\small \text{(ii) system power balance}}\\[-2.5pt]
 & {\small \text{(iii) load and generator limits,}}
\end{align*}
where $c_{{\rm gen}}$ is the vector of generation marginal costs, and $c_{{\rm load}}$ is the vectorized marginal prices at which loads value delivered power. In the day-ahead market, electricity is offered and sold for the 24 hours of the next day in (hourly) blocks. The electricity price and volume for each hour are determined at the intersection of demand and supply, typically at around noon. Within the context of \eqref{eq: zms}, renewable curtailment can sometimes alleviate line flow congestion, thus allowing for an increase in the amount of load served by cheap fossil-fuel generation and raising the market surplus.

\begin{figure}
\centering
\includegraphics[width=0.975\linewidth]{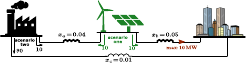}
\caption{Congested power system. Due to impedance ratios and line flow limitations, the green generator can only provide 20 MW to the load (with no black) in Scenario 1, while the black generator can provide 100 MW to the load (with no green) in Scenario 2. Every unit of green, therefore, can be replaced with 5 units of black. If the load values power highly enough, the market surplus will be maximized by fully curtailing the zero-marginal-cost renewables and maximizing black generation.}
\label{fig:congestion}
\end{figure}

\textit{\textbf{Congestion Example:}} A very simple example of this phenomenon is illustrated in Fig.~\ref{fig:congestion}, where an operator clears this market by solving \eqref{eq: zms}.
In this network, ``black'' (i.e., fossil fuel) generation sees two electrical paths to the load: a low impedance path (line $\boldsymbol c$), and a high impedance path (lines $\boldsymbol a$ and $\boldsymbol b$). Meanwhile, ``green'' (i.e., renewable) generation sees two high impedance paths to the load. There is a 10 MW flow constraint on line $\boldsymbol b$. We consider two scenarios:
\begin{itemize}
    \item Scenario 1: If black produces nothing, green can only provide 20 MW to the load, since 10 MW will flow on every line due to impedance ratios and flow constraints (Scenario 1).
    \item Scenario 2: Assume green turns down production by 2 MW, yielding 1 MW of capacity on the constrained line. If black then increases its production, it will send power in both directions. Due to impedance ratios, a current divider~\cite{nilsson2020electric} shows the percentage of power sent in these directions:
    \begin{subequations}
    \begin{align}
    \boldsymbol{a}{\text-}\boldsymbol{b}\;{\rm flow} & =\tfrac{0.01\,\Omega}{0.01\,\Omega+0.09\,\Omega}=10\%\\
    \boldsymbol{c}\;{\rm flow} & =\tfrac{0.09\,\Omega}{0.01\,\Omega+0.09\,\Omega}=90\%.
    \end{align}
    \end{subequations}
    Thus, if black produces 10 MW, 1 MW will flow on lines $\boldsymbol a$-$\boldsymbol b$, and \textit{9 MW} will flow on on line $\boldsymbol c$.
\end{itemize}
Ultimately, green can independently deliver $10+10=20$ MW, while black can independently deliver $10+90=100$ MW. And this is solely due to the structure of the transmission grid and the location where each generator connects to it. According to \eqref{eq: zms}, as long as the delivered power is 25\% more valuable to the consumer than its cost of production, the renewables will be shut off entirely\footnote{According to \eqref{eq: zms}, black power will lead to larger market surplus if $c_{{\rm load}}\times100\,{\rm MW}-c_{{\rm gen,black}}\times100\,{\rm MW}>c_{{\rm load}}\times20\,{\rm MW}-0\times20\,{\rm MW}$, assuming renewables have an offer cost of $0$. This inequality simplifies to $c_{{\rm load}}>1.25c_{{\rm gen,black}}$.}.{\hspace*{0em plus 1fill}\makebox{$\quad\qedsymbol{}$}} 

From a market clearing context, renewable curtailment is a complex function of meshed network topologies, impedance ratios, flow constraints, and marginal pricing valuations. The example in Fig.~\ref{fig:congestion} in particular can be considered an example of Braess' paradox~\cite{blumsack2006braess,Crozier:2022}: the removal of line $\boldsymbol b$ could actually alleviate congestion, allowing green and black to produce at their full capacities. Regardless, it is the \textit{design} of \eqref{eq: zms} which leads to renewable curtailment, and this design failure is rooted in the anachronistic assumption that power produced by green and black sources is 1:1 interchangeable. In contrast, we propose market clearing and marginal pricing strategies around the concept of \textit{one-way interchangeability}, i.e., appetite for electrical power can be satisfied by green \textit{or} black generation, but appetite for green power can only be satisfied by green generation.

\subsection*{Paper outlook and contributions} Our central claim is this: the market surplus functions which have governed deregulated electricity markets for over three decades no longer truly maximize social welfare, as they do not accurately account for the societal value of green energy. Despite policy makers' attempts to cap emissions and subsidize renewable energy, curtailment of renewable resources worldwide due to negative price exposure and exclusionary dispatch~\cite{vt_curtail} is evidence enough that updates to the market clearing process are needed. Allowing consumers to directly bid their preference for renewable power can provide a market-based solution to this problem and internalizes the externality of energy source preference. We argue that inclusion of such a bidding and pricing strategy into competitive market clearing algorithms can provide the following benefits to congested power systems:
\begin{enumerate}[label=(\roman*)]\itemsep-0.25em 
    \item dispatch solutions which prioritize renewables, 
    \item nodal pricing signals which spur renewable investments, 
    \item and a more effective distribution of green energy benefits.
\end{enumerate}




To achieve these ends, we first develop a day-ahead market clearing function which allows consumers to tractably bid their preference for renewable energy within the dual bids they submit to the market clearing operator. Next, based on the optimal dispatch, we devise a nodal pricing strategy which broadcasts real-time pricing signals and efficiently remunerates all system generators. Finally, we study the deployment of these algorithms within small toy power grids (to provide intuition) and a synthetic replica of the Texas (ERCOT) grid.


\section*{Dual Pricing Dispatch} 
\subsection*{Willingness to pay}
As environmental awareness increases, consumers are growingly inclined to support sustainable practices and make environmentally conscious choices. This reflects into a general willingness to pay (WTP) for green energy. Research shows that consumers are indeed willing to pay a premium for green energy, although preferences may vary across regions and among individuals, contingent upon factors such as renewable energy technologies and local contexts ~\cite{SUNDT20151, LONGO2008140, CERDA2024107301}. 

To meet this rising demand for sustainability, we define a market mechanism which enables load serving entities to directly influence their allocation of energy by bidding their WTP for green energy into the clearing market. By allowing these entities to prioritize green energy sources, such as wind, hydro, solar and nuclear power\footnote{The designation of what constitutes green power, black power, or anything in between is beyond the scope of this paper. We focus on designing a market mechanism for their inclusion in the bidding process.}, over conventional sources like natural gas and coal, our proposed mechanism not only meets the energy requirements of load serving entities, but it additionally fulfills their potential preference for environmentally responsible energy consumption. If consumers are willing to pay more, this framework allows the maximization of green energy by strategically dispatching black energy to support the flow of green energy. In this approach, load serving entities receive energy from both categories of generation (i.e., renewable energy sources versus conventional sources) separately, and consumers pay an additional amount for the proportional quantity of renewable energy they receive. 


\subsection*{Market clearing function} 
\begin{figure}[h]
    \centering \includegraphics[width=0.4\textwidth]{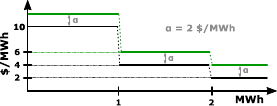}
    \caption{The multi-part bid of a single load is depicted. In this bid, the load values the first unit of energy at \$10/MWh, the second unit of energy at \$4/MWh, and so on. The $\alpha=2$ value bid by this load indicates that the load is willing to pay an additional \$2/MWh, at any load interval, for energy which comes from a green source. Constant $\alpha$ values across a given multi-part bid are essential for preserving problem convexity.}
    \label{fig:alpha}
\end{figure}

Consumers' WTP more for green energy is included in the day-ahead market clearing function by a vector $\alpha$ (\$/MWh):
\begin{align*}
\max_{p_{{\rm load}},p_{{\rm gen}}}\quad & c_{{\rm load}}^{T}p_{{\rm load}}-c_{{\rm gen}}^{T}p_{{\rm gen}}+\alpha^T p^{{(\textcolor{nicegreen}{\bf g})}}_{\rm load}
\numberthis\label{eq: dpd}\\
{\rm s.t.}\quad\;\;\; & \small{\text{(i) line flow constraints}}\\[-2.5pt]
 & \small{\text{(ii) system power balance}} \\[-2.5pt]
 & \small{\text{(iii) load and generator limits}} \\[-2.5pt] 
 & \small{\text{(iv) \textbf{green power balance}}.} 
\end{align*}
In \eqref{eq: dpd}, $c_{{\rm load}}^{T}$ and $c_{{\rm gen}}^{T}$ represent standard, multi-part bid and offer curves, but $\alpha$ reflects an additional bid submitted by load serving entities which reflects their WTP for renewable sources over fossil fuel sources; $p^{{(\textcolor{nicegreen}{\bf g})}}_{\rm load}$, meanwhile, is the amount of green energy proportioned to the corresponding load. The scalar $\alpha$ value, associated with a single multi-part load bid, is depicted in~\ref{fig:alpha}. Fundamentally, $\alpha$ gives load bidders the ability to financially quantify the relationship between their consumer welfare, and the amount of green power they receive to power their load.

The green power balance constraint in \eqref{eq: dpd} assigns dispatched green energy to consumers in accordance with this WTP. Despite the additional green energy balance constraint and WTP term in the objective, this updated social welfare maximization problem is still a convex linear program (LP). The full formulation of the problem is presented in the SI.

This formulation maximizes the social welfare by including the environmental preferences of consumers. The increase of social welfare in \eqref{eq: dpd}, compared to \eqref{eq: zms}, is proportional to the amount of green energy that the load serving entities receive and the monetary value that consumers place on renewable energy. The more consumers are willing to pay for green energy, the higher the social welfare. In cases where consumers do not demonstrate a preference for green energy (i.e., $\alpha = 0$), the market operates according to the traditional mechanism outlined in \eqref{eq: zms} (i.e., \eqref{eq: dpd} would revert to \eqref{eq: zms}). In the formulation, $\alpha$ bids submitted by load serving entities must be nonnegative, i.e., $\alpha\ge 0$, 
where a bid of $\alpha=0$ means indifference to the origin of the energy source, and a bid higher than $\alpha=0$ indicates a preference for green energy over black energy (see the SI for details). In the face of congestion, positive $\alpha$ values monotonically raise the clearing prices and increases the dispatch of green energy, when possible. This framework thus reflects the market's desires, maximizing green energy dispatch when $\alpha > 0$.

\subsection*{Black and green LMPs} 
Due to the structure of our framework, we can derive both black and green LMPs (i.e., the marginal cost of serving the next unit of energy, vs the marginal cost of serving the next using of \textit{green} energy, at a given node). The nodal vector of green LMPs, denoted ${\rm LMP_{\textcolor{nicegreen}{\bf g}}}$, is equal to the nodal vector of black LMPs, $\rm LMP_{\bf b}$, plus a scalar offset $\lambda_{\textcolor{nicegreen}{g}}$, which is the dual variable (i.e., shadow price) associated with the green power balance constraint in \eqref{eq: dpd}:
\begin{align} \label{eq:lambdagreen}
{\rm LMP_{\textcolor{nicegreen}{\bf g}} = \rm LMP_{\bf b} + \lambda_{\textcolor{nicegreen}{g}}}.
\end{align}
These LMPs are derived in the SI. The green power balance constraint stems from the fundamental principle that the total amount of green power that the loads are apportioned should be equal to the total power that is provided by the green generators in the system. Thus, $\lambda_{\textcolor{nicegreen}{g}}$ represents the marginal extra cost of supplying an additional unit of green electricity (e.g., by shifting the congested generation mix to allow more green power to flow). These LMPs are used to proportionally charge power consumers and remunerate power producers. Specifically, 
\begin{itemize}
\item Consumers/producers of energy designated as green pay/are paid at the marginal rate of ${\rm LMP_{\textcolor{nicegreen}{\bf g}}}$, 
\item Consumers/producers of energy \textbf{not} designated as green pay/are paid at the marginal rate of $\rm LMP_{\bf b}$, 
\end{itemize}

These LMPs consequently give signals to where green producers should build generators, as higher green LMPs will generate higher profits. Furthermore, loads interested in consuming green power should locate to regions where green LMPs are correspondingly low.

\subsection*{Motivating example} Let us illustrate this market-clearing procedure using the following examples where congestion occurs. We consider a one hour snapshot of operation, and we note that these examples are relevant for both nodal (where nodes represent physical system buses) and zonal (where nodes represent aggregated regions) power system markets.

\begin{figure}[h]
    \centering \includegraphics[width=0.3\textwidth]{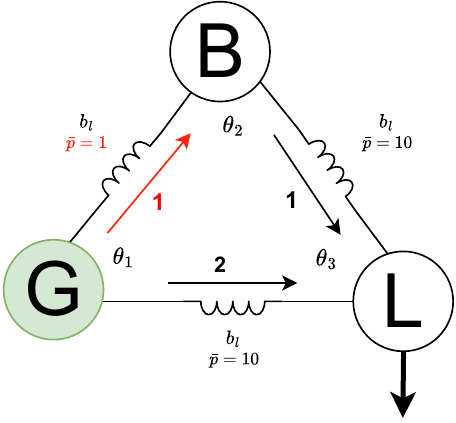}
    \caption{The power flows in a 3-node network containing a renewable generator (G), a black (B) generation source, and a load (L).}
    \label{fig:3bussystem_flows}
\end{figure}

Fig. \ref{fig:3bussystem_flows} shows a network that has 3 nodes (buses). This network consists of a green generator with a capacity of 4MW, a black generator with a capacity of 4MW and a load with an elastic demand. The impedances of all connected lines are identical. The green generator offers their production at a marginal price of \$0/MWh, while the black generator submits a marginal offer of \$10/MWh. The load is willing to pay \$4/MWh.

\begin{figure}[h]
    \centering
    \includegraphics[width=0.35\textwidth]{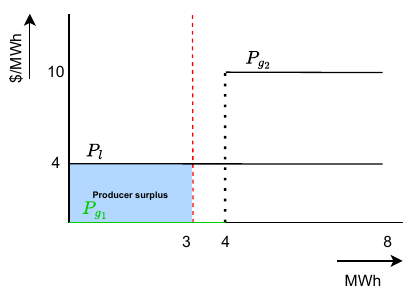}
    \caption{The bids of the generators and the loads}
    \label{fig:3bussystem_demand}
\end{figure}

If the dispatch is cleared via standard social welfare maximization (i.e., \eqref{eq: zms}), the green generator would dispatch 3 MWh of energy, and the load would receive 3 MWh of green energy. According to Fig. \ref{fig:3bussystem_demand}, however, the intersection of the supply and demand curves is 4 MWh, indicating that while the load is willing to pay for an additional unit of green power, the congested network cannot support the dispatch of an additional unit from the green generator. 

\begin{figure}[h]
    \centering \includegraphics[width=0.3\textwidth]{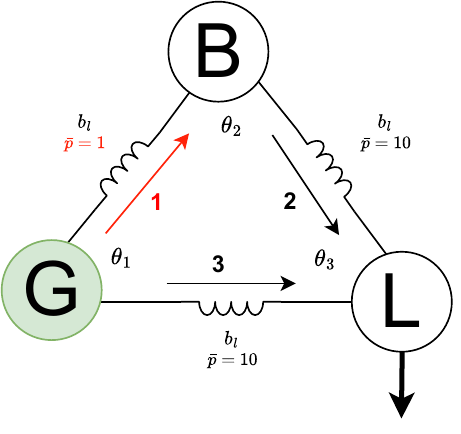}
    \caption{The power flows in a 3-node network using the dual pricing dispatch framework}
    \label{fig:3bussystem_DPD}
\end{figure}

\begin{figure}[h]
    \centering \includegraphics[width=0.35\textwidth]{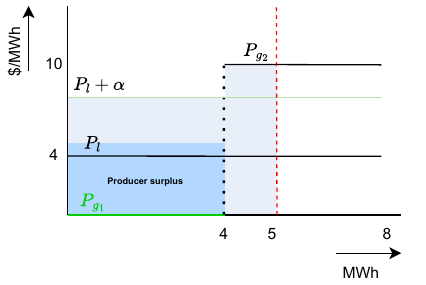}
    \caption{The bids of the generators and the loads using the dual pricing dispatch framework}
    \label{fig:3bussystem_DPD_bids}
\end{figure}

\textbf{Dual dispatch:} We now apply the dual dispatch clearing framework \eqref{eq: dpd} to the same example. In this example, we assume the load is willing to pay $\alpha=$\$3/MWh extra for green energy, thus internalizing the externality of energy source preference. As depicted in Fig~\ref{fig:3bussystem_DPD_bids}, the load receives an additional unit of green at the expense of 1 MWh dispatched by the black generator, for a total of 5 MWh. In effect, the 1 MWh of dispatched black alleviates congestion, allowing one more MWh of green energy to flow, as shown in Fig.~\ref{fig:3bussystem_DPD}.


\begin{figure}[h]
    \centering \includegraphics[width=0.3\textwidth]{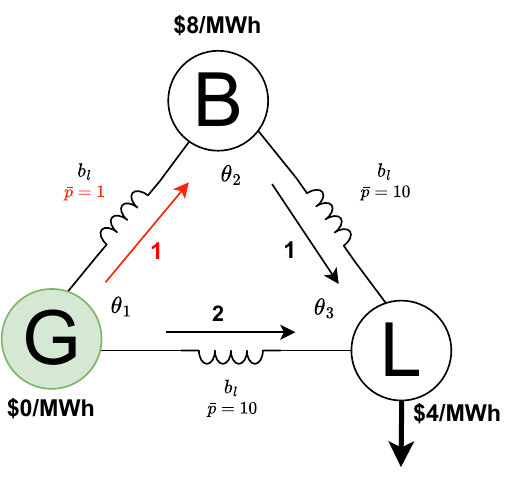}
    \caption{The LMPs in the 3-node network}
    \label{fig:3bussystemLMPS}
\end{figure}

Now, we will consider the effect of $\alpha$ on the clearing prices in a congested system. The LMPs are derived from the infinitesimal sensitivity of the system to changes in power supply and demand. The load pays the LMP at the connected node and the generators receive the LMP at the node where the energy is generated. The mathematical derivation of LMPs can be found in the SI. In the absence of any network constraints, the price of supplying an additional unit of power would be equal at all 3 nodes. In this congested network, the default dispatch gives the following varying LMPs for the 3 nodes as illustrated in Fig. \ref{fig:3bussystemLMPS}. 

While the cost of supplying an additional infinitesimal unit of power at the green generator is \$0/MWh (as the generator is not running at full capacity yet), the cost of supplying an extra infinitesimal unit of power at the black generator would be \$8/MWh. 

\begin{figure}[h]
    \centering \includegraphics[width=0.3\textwidth]{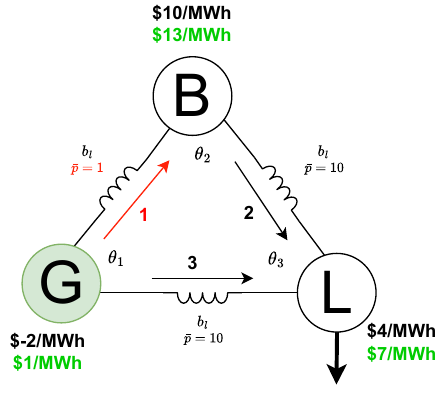}
    \caption{The black and green LMPs in the 3-node network}
    \label{fig:3bussystem_blackgreenLMPs}
\end{figure}

In Fig. \ref{fig:3bussystem_blackgreenLMPs}, we distinguish between black and green LMPs in the case that $\alpha$ is equal to \$3/MWh. 
This figure shows that the green LMPs are monotonically higher than the black LMPs by $\alpha$. Although for larger systems it does not generally hold that $\lambda_{\textcolor{nicegreen}{g}}$ is equal to $\alpha$, as multiple loads will bid different $\alpha$'s, the green LMPs will always be higher than the black LMPs (if the load values green energy more than black energy).

At the generator nodes, the LMPs have now changed compared to Fig. \ref{fig:3bussystemLMPS}. In order to serve a small increment of load at the black generator, the black generator would have to dispatch some additional energy, and the cost would be \$10/MWh. 

\section*{Dual Pricing Dispatch in Texas} 
\subsection*{The case of Texas}To show the framework on a larger scale, we demonstrate the framework on the Texas synthetic test case,  as designed by Texas A\&M University ~\cite{texasamtestcases, Birchfield_testcase, xuhics, xuirep, kunkolienkar}. This is an open-source network which resembles the real Texas grid in structure and complexity. It consists of 2000 nodes and includes the costs of the generators.
In the base case scenario of the synthetic Texas power system, 18.7\% of the total generation capacity is renewable energy and 82.3\% is covered by other sources, i.e., coal and natural gas. As more countries, organizations and companies aim to reach net-zero by 2050, Texas has considerable potential for expanding its renewable energy capacity \cite{usenergyinformation}. In light of these decarbonization efforts, power systems will be increasingly dominated by renewable energy sources. 
Projections show that total generation in the ERCOT system will consist of 53\% combined wind and solar generation by 2035 \cite{USEIA}, and that transmission system limits will cause most curtailments in the summer. Congestion costs in Texas may rise to \$2.8 billion in 2035 \cite{USEIA}. To mimic this scenario, we modify the ERCOT system such that 50\% of the generation capacity is covered by renewable energy sources.

\subsection*{Day-Ahead Market Clearing} In the day-ahead market, offers from the generators are ranked from lower prices to higher prices. This aggregated curve is referred to as the merit-order curve (the supply curve). In the absence of network constraints, the intersection of the aggregated demand with the supply curve determines the clearing price and clearing volume. The clearing price corresponds to the marginal cost of the most expensive generator dispatched. All participating generators receive this price for their units, regardless of their supply bid price. This principle applies to buyers as well. Regardless of their willingness to pay, all consumers purchasing electricity within the market pay the same marginal price determined by the intersection of supply and demand. 

\begin{figure}[h]
    \centering \includegraphics[width=0.45\textwidth]{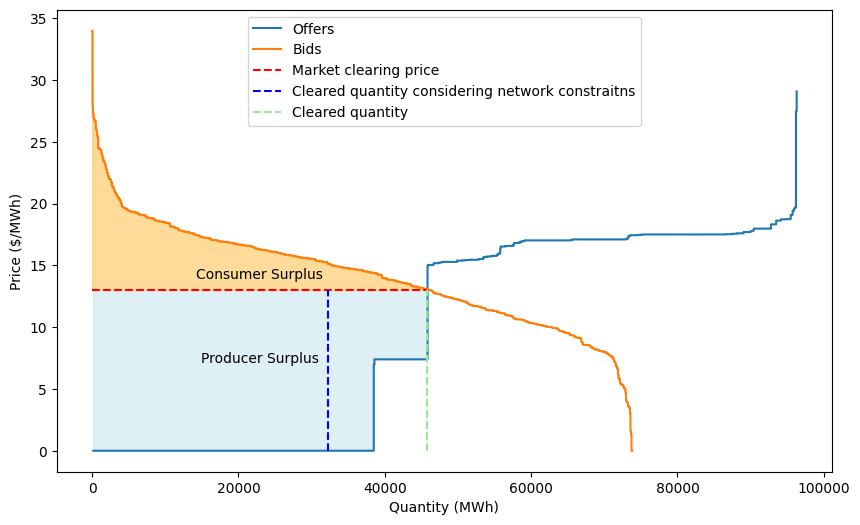}
    \caption{The aggregated demand curve and merit order curve for one hour}
    \label{fig:bidsercot}
\end{figure}

\begin{figure}[h]
    \centering \includegraphics[width=0.5\textwidth]{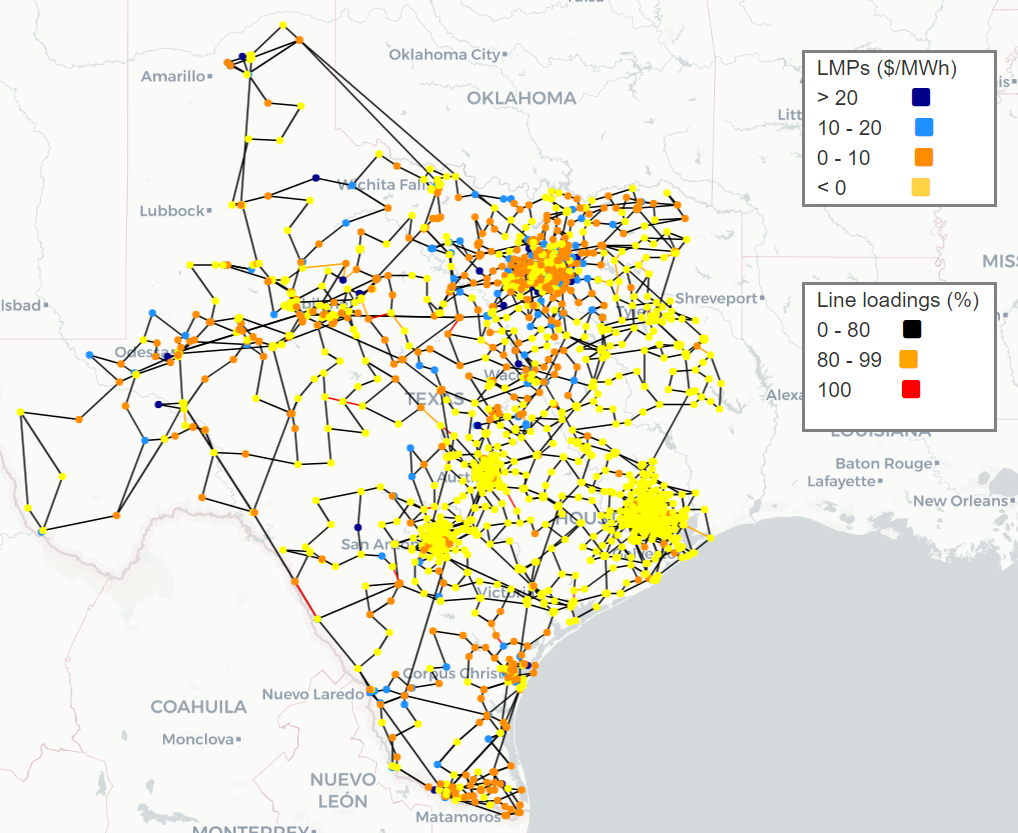}
    \caption{LMPs in Texas with 50\% RES}
    \label{fig:texas_blackLMPs_50RES}
\end{figure}

Fig. \ref{fig:bidsercot} shows the aggregated demand and supply curves for the ERCOT system for one hour, where the market would be cleared at \$12.16/MWh, for a volume of 45,809 MWh, without the dual dispatch framework and when we disregard transmission constraints (i.e., there is no congestion considered). At this clearing price, 100\% of the buyers are supplied with green electricity. This dispatch allows 95\% of the green generators participating in the market to produce at their full capacity. 
If we now consider transmission line constraints, the total energy that is dispatched is 32,191 MWh. The market is no longer cleared at the same equilibrium.
The green generators are no longer fully dispatched. Now, instead of 94.8\%, 66.7\% of the total green generation is dispatched due to network constraints, even though all green generators have 0 or low marginal costs. The distribution of the LMPs per node in Texas is shown in Fig. \ref{fig:texas_blackLMPs_50RES}, highlighting a substantial number of nodes with negative LMPs. The network has 38 lines that are fully congested.

\subsection*{Day-Ahead Market Clearing using Dual Pricing Dispatch} 
\begin{figure}[h]
    \centering \includegraphics[width=0.35\textwidth]{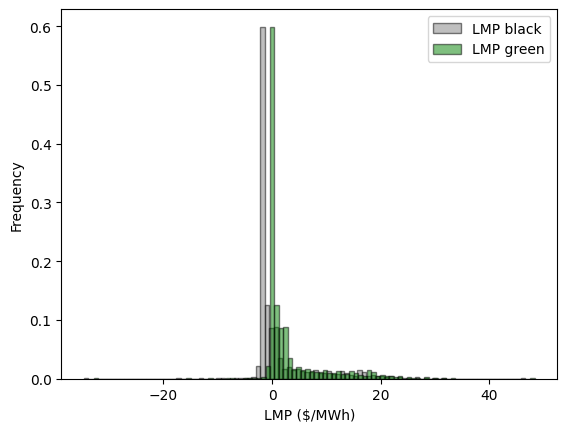}
    \caption{Black and green LMPs in Texas with 50\% RES}
    \label{fig:texasblack_greenLMPs_50RES}
\end{figure}

\begin{figure}[h]
    \centering \includegraphics[width=0.5\textwidth]{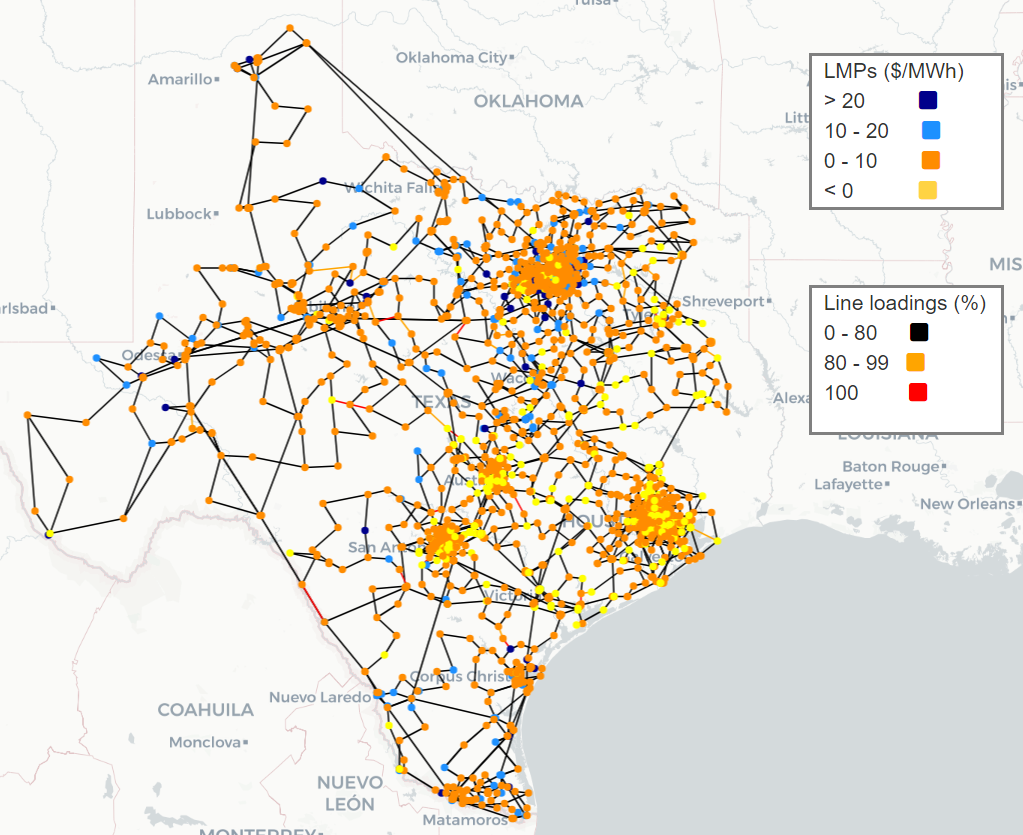}
    \caption{Green LMPs in Texas with 50\% RES}
    \label{fig:texas_greenLMPs_50RES}
\end{figure}

Next, we apply the dual pricing dispatch concept with the aim to let consumers bid their preferences for green energy, alleviate congestion and dispatch more green energy. The bids of the consumers for renewable energy are normally distributed around \$5/MWh with a standard deviation of \$1/MWh. This allows the dispatch of an extra 239 MWh of green electricity, while black energy is increased with 8 MWh. If we assume that one megawatt can supply 800 homes on a normal day in Texas \cite{powerdemandtexas}, 239 MWh can power 191,200 extra homes for one hour with green energy. The black and green LMPs in the dual dispatch framework are shown in Fig. \ref{fig:texasblack_greenLMPs_50RES}, indicating that the green LMPs are generally around \$2 higher from the black LMPs, as $\lambda_{\textcolor{nicegreen}{\textbf{g}}}=1.78$ \$/MWh.  The LMPs are centered around \$0/MWh, as the system has a large share of renewable energy sources with 0 or low marginal costs. How the black and green LMPs differ across Texas is shown in Fig. \ref{fig:texas_blackLMPs_50RES} and \ref{fig:texas_greenLMPs_50RES} respectively. The map of the green LMPs demonstrates more positive LMPs than the map of the black LMPs, where a major part of the LMPs is negative. We now observe 36 congested lines in the network. 

\subsection*{Varying renewable energy capacity} 
In this section, we will examine a system with varying levels of renewable energy as compared to the previous scenario where the energy supply is equally divided between 50\% renewable and 50\% non-renewable sources. First, we will examine the initial scenario of the synthetic test case where 18.7\% of the total generation capacity comes from renewable energy, and then we will gradually increase the amount of RES in the system to 30\%, 60\%, 70\% and 80\%. An overview of the additional black and green energy dispatch for the different RES scenarios is given in Table \ref{tab:additional_dispatch}.

\begin{table}[h!]
\centering
\caption{Additional Dispatch of Green and Black Energy}
\begin{tabular}{lcc}
\toprule
\textbf{RES Percentage} & \textbf{Green Energy (MWh)} & \textbf{Black Energy (MWh} \\
\midrule
18.7\% RES & 10.7 & 2060.9 \\
30\% RES & 13.2 & 22.2 \\
50\% RES & 239.0 & 8.0 \\
60\% RES & 297.3 & 3.7 \\
70\% RES & 300.6 & 0 \\
80\% RES & 249.0 & 0 \\
\bottomrule
\end{tabular}
\label{tab:additional_dispatch}
\end{table}

\begin{figure}[h]
    \centering \includegraphics[width=0.35\textwidth]{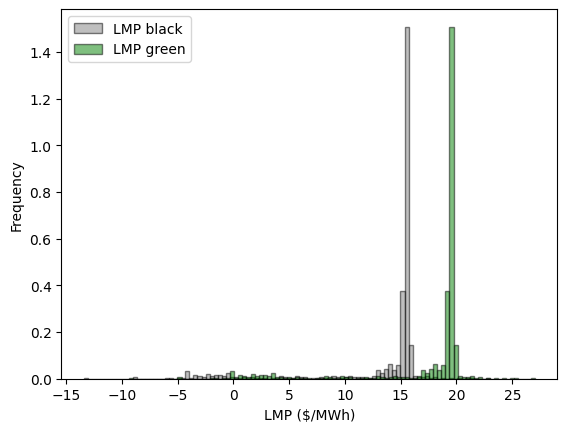}
    \caption{Black and green LMPs in Texas with 18.7\% RES}
    \label{fig:texasblack_greenLMPs_case1}
\end{figure}

Solving the optimization problem which maximizes social welfare for the 18.7\% RES system allows 94.3\% of the entire load (17,953 MWh) to be covered by renewable energy sources, with 5.7\% black energy is dispatched. If consumers value green electricity over black energy as in the previous case, the dispatch of black sources will increase to 2061 MWh, which causes an increase of around 11 MWh of green electricity in the grid. 
For this system, the green LMPs in this case are generally around 4 \$/MWh higher than the black LMPs, as is shown in Fig. \ref{fig:texasblack_greenLMPs_case1}.

Let us now expand the RES to 30\%. Here, the extra black energy dispatched is much less than for the 18.7\% RES case, while the extra green energy dispatch increases significantly. Moving to systems with larger shares of RES, more green energy is integrated in the grid, while less black energy has to be dispatched to squeeze out additional green energy until we reach 80\% RES. For the 70\% and 80\% RES cases, no additional black energy is dispatched, however, there is an increase in green energy when we apply the dual dispatch framework. This is due to the fact that not all generators classified as `green', offer their energy at \$0 marginal cost. For example nuclear energy has higher marginal costs, but is classified as `green' in this paper. 

For each case, we investigated how the $\lambda_{\textcolor{nicegreen}{g}}$ changes for each scenario, to see the rise in green LMPs compared to black LMPs, see \eqref{eq:lambdagreen}. Table \ref{tab:lambdagreen} compares the increase in green LMPs with respect to the black LMPs. 

\begin{table}
\centering 
\caption{Comparison of $\lambda_{\textcolor{nicegreen}{\textbf{g}}}$ for different cases}
\begin{tabular}{lrrrrrr} \label{tab:lambdagreen}
\textbf{RES} & 18.7\% & 30\% & 50\% & 60\% & 70\% & 80\% \\
\midrule
$\lambda_{\textcolor{nicegreen}{g}}$ (\$/MWh) & 4.03 & 1.78 & 1.78 & 1.78 & 1.14 & 0.67 \\
\bottomrule
\end{tabular}
\end{table}

\begin{figure}[h]
    \centering \includegraphics[width=0.35\textwidth]{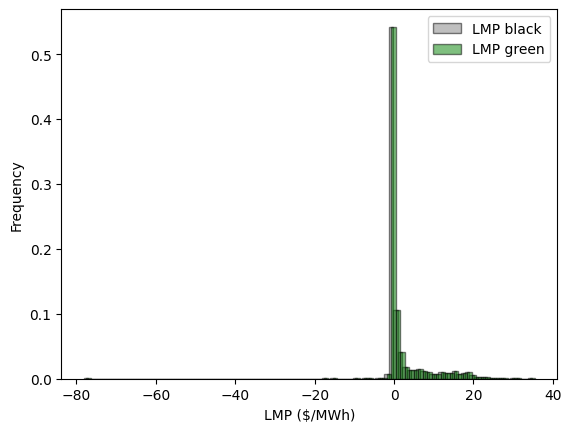}
    \caption{Black and green LMPs in Texas with 80\% RES}
    \label{fig:texasblack_greenLMPs_80RES}
\end{figure}

The more RES are present in the system, the smaller is $\lambda_{\textcolor{nicegreen}{g}}$ and the smaller is the difference between green and black LMPs. In the case of 80\% RES, the green LMPs are only $\lambda_{\textcolor{nicegreen}{g}}=0.67$ \$/MWh higher than the black LMPs, which is shown in Fig. \ref{fig:texasblack_greenLMPs_80RES}. 

Although the $\lambda_{\textcolor{nicegreen}{g}}$ is small, this does not mean that there is a smaller impact on the green energy dispatch. For example, the application of the framework to the scenario with 60\% of RES in the system with only a minimal increase in the black energy dispatch, leads to a much larger increase in RES than in the case, where there is only 18.7\% RES. The extra dispatch of green energy is dependent on the grid topology and distribution of RES in the grid. Table \ref{tab:additional_dispatch} illustrates clearly that higher penetration of RES in the grid leads to larger increases in green energy dispatch with just a marginal increase in black energy. However, only in the initial case with 18.7\% RES the additional green energy dispatch is minimal compared to the extra black energy dispatch. 



\subsection*{Impact on carbon emissions}  
If the buyers in the market would like to use as much of the available renewable energy as possible, the framework supports the flow of green energy by dispatching black energy. Black energy (coal and natural gas) emissions are generally much higher (380-1050 kg/MWh) than green emissions (2-190 kg/MWh) \cite{TURCONI2013555} and it might seem counter-intuitive to incorporate more black energy. However, the goal of the framework is to add more green energy to the grid, potentially lowering the average CO2e emissions per MWh. The carbon emission impact depends on the ratio between the additional dispatch of green and black energy. Considering the scenarios where RES $>$ 50\%, the average carbon emissions are slightly reduced as illustrated in Table \ref{tab:emissions}.

\begin{table}[tbhp]
\centering
\caption{Average CO2e Emissions (kg/MWh)}
\begin{tabular}{lrr}
\toprule
& \multicolumn{2}{c}{Average emissions (kg/MWh)} \\
\cmidrule{2-3}
            & Without DPD & With DPD \\
\midrule
18.7\% RES  & 125.9       & 142.3    \\
30\% RES    & 109.1       & 109.6    \\
50\% RES    & 142.5       & 142.6    \\
60\% RES    & 142.5       & 142.2    \\
70\% RES    & 142.5       & 142.1    \\
80\% RES    & 142.5       & 142.1    \\
\bottomrule
\end{tabular}
\label{tab:emissions}
\end{table}

For reference, in June 2024, the average carbon emissions in Texas were 367 kg/MWh \cite{emissionsTexas}, mostly due to the dispatch of natural gas, responsible for 65\% of the carbon emissions. \\ 

The operation of black energy sources is necessary to allow for more green energy integration, although these black sources generally have much higher emissions than green sources. With the proposed market framework, we derive a price for the dispatch of extra green energy and extra black energy; $\lambda_{\textcolor{nicegreen}{\textbf{g}}}$, if consumers are willing to pay more for green energy. This can be seen as
the economic and societal value of the dispatch of the adding more renewable energy to the grid. By distinguishing between green and black energy prices, we can even derive a carbon taxation system, where the end-users pay the price of the CO2e emissions, as they now have the choice of being supplied with green or black energy, based on their $\alpha$-bid.

\section*{Concluding Remarks}
Efforts to promote the integration of renewable energy have mainly been driven by external mechanisms and incentives. In this work, we introduce a novel market-clearing function that internalizes consumers' willingness to pay for green energy directly within the market framework. By embedding consumer preferences for renewable energy into the market-clearing function, we establish a transparent framework that allows consumers to express their preference for renewable energy sources. This approach optimizes the dispatch of renewable energy, while respecting the flow constraints of the power system. Furthermore, it generates market signals for energy providers, incentivizing greater investment in renewable energy sources.

The proposed market clearing function obtains, from an economic point of view, efficient market clearing prices, which allows more green energy to be consumed. We distinguish between black and green locational marginal prices, where the green locational marginal prices are generally higher than black locational marginal price under the assumption that some consumers are inclined to pay extra for green energy. The approach has been tested on the synthetic Texas case with 2000 nodes for different scenarios. For systems with a high RES percentage (more than 50\%), the additional green energy dispatch is in the range of 239-300 MWh, which can power up to 240,000 additional homes. In addition, our market-clearing function lowers the average carbon emissions per MWh.

With this work, we would like to encourage policy makers to incorporate consumer preferences in the market clearing function, not only to give consumers influence in choosing their electricity source, but also to alleviate curtailment of renewable energy sources in congested networks.
Future work involves expanding this market clearing function to a market design.

\acknow{Please include your acknowledgments here, set in a single paragraph. Please do not include any acknowledgments in the Supporting Information, or anywhere else in the manuscript.} 


\bibliography{references}
\newpage
\onecolumn



\definecolor{nicegreen}{rgb}{0, 0.593, 0}

\section*{Appendix Supporting Information}






\subsection*{Full formulation of the market clearing algorithm}

In the optimal power flow (OPF) problem, loads are fixed. However, in a market clearing problem, loads elastically bid for power. The market clearing program utilized in this paper can be formulated as
\begin{subequations}\label{eq:clearing}
\begin{align} \label{eq:dammarketclearing}
\max_{p_{{\rm load}},p_{{\rm gen}}}\quad & \sum_{i\in\mathcal{L}}\sum_{j\in\mathcal{B}_{i}}\left(p_{{\rm load},i}^{(j)}\times c_{{\rm load},i}^{(j)}\right)-\sum_{k\in\mathcal{G}}\sum_{l\in\mathcal{O}_{k}}\left(p_{{\rm gen},k}^{(l)}\times c_{{\rm gen},k}^{(l)}\right)\\
{\rm s.t.}\quad & \underline{p}_{{\rm load},i}^{(j)} \le p_{{\rm load},i}^{(j)}\le\overline{p}_{{\rm load},i}^{(j)},\quad\forall i\in\mathcal{L},\forall j\in\mathcal{B}_{i} \label{eq:subeqloadlimits}\\
 & \underline{p}_{{\rm gen},k}^{(l)}\le p_{{\rm gen},k}^{(l)}\le\overline{p}_{{\rm gen},k}^{(l)},\quad\forall k\in\mathcal{G},\forall l\in\mathcal{O}_{k} \label{eq:subeqgenlimits}\\
 & \sum_{k\in\mathcal{G}_{n}}\sum_{l\in\mathcal{O}_{k}}p_{{\rm gen},k}^{(l)}-\sum_{i\in\mathcal{L}_{n}}\sum_{j\in\mathcal{B}_{i}}p_{{\rm load},i}^{(j)}=\sum_{m\in\Lambda_{n}}B_{nm}(\delta_{n}-\delta_{m}),\quad\forall n\in\mathcal{N}:\;\lambda_{n} \label{eq:subeq2}\\
 & \underline{p}_{nm}\le B_{nm}(\delta_{n}-\delta_{m})\le\overline{p}_{nm},\label{eq:subeqflowlimits}
\end{align}
\end{subequations}
where $\mathcal{L}$ is the set of all loads, $\mathcal{L}_{n}$ is the set of all load at bus $n$, $\mathcal{B}_{i}$ is the set of bid blocks submitted by a load, $\mathcal{G}$ is the set of generators, $\mathcal{G}_n$ is the set of generators at bus $n$, $\mathcal{O}_{k}$ is the set of generator $k$ offer blocks, and $\Lambda_{n}$ is the set of buses $m$ connected to bus $n$ (i.e., $m\in\Lambda_{n}$). Furthermore, $B_{nm}$ is the susceptance of the line between buses $n$ and $m$ (whose flow limits are given as $\underline{p}_{nm}$, $\overline{p}_{nm}$), while $\delta_n$ and $\delta_m$ are the associated nodal phase angles. In \eqref{eq:clearing}, the limits of the $j^{\rm th}$ bid block of the $i^{\rm th}$ load (at valuation $c_{{\rm load},i}^{(j)}$) are captured by \eqref{eq:subeqloadlimits}, and the limits of the  $l^{\rm th}$ offer block of the $k^{\rm th}$ generator (at cost $c_{{\rm gen},k}^{(l)}$) are captured by \eqref{eq:subeqgenlimits}. Active power balance is captured by \eqref{eq:subeq2}, and transmission line constraints are captured by \eqref{eq:subeqflowlimits}. Each of the $n$ power balance equations has an associated dual variable, given as $\lambda_n$, which represents the shadow price associated with a load perturbation; this is equal to the nodal clearing price. This formulation does not take transmission line losses into account, and it is based on the DC-OPF formulation.

\subsection*{Dual Pricing Dispatch Market Clearing Algorithm}
\begin{figure}[h]
    \centering \includegraphics[width=0.7\textwidth]{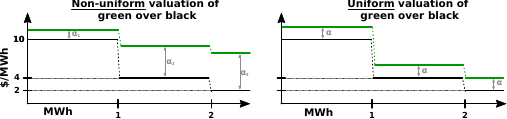}
    \caption{Load bids with non-uniform (left) and uniform (right) $\alpha$ values.}
    \label{fig:alpha_vs}
\end{figure}

Allowing consumers to bid extra money for energy served by green sources is a nontrivial alteration to the market clearing problem. Loads and generators submit bids and offers in ``blocks''; Fig.~\ref{fig:alpha_vs}, for example, shows the standard block bid submitted by a load. In this figure, we also show the addition of $\alpha$ parameters, which correspond to a load's  willingness to pay a premium for energy served by green, instead of black, energy sources. In the left sub-figure, $\alpha$ is non-uniform across the bid blocks. Allowing an optimizer to dispatch both green and black energy to serve a given load without selecting overlapping bid-blocks requires nonlinear constraints, which engenders non-convexity. To overcome this problem, we allow loads to bid a uniform $\alpha$ value, where the increase in value of green over black is uniform across all bid blocks (for a given load). This allows us to write the updated market clearing problem, where loads bid their preference for green, linearly.

To formulate the updated market clearing problem, we first split the generation into two sets: green generators $\mathcal{G}^{(\textcolor{nicegreen}{{\bf g}})}$ and black generators $\mathcal{G}^{({\bf b})}$. We also split the energy served to a given load $p_{{\rm load},i}$ into two constituent portions: green energy $p_{{\rm load},i}^{(\textcolor{nicegreen}{{\bf g}})}$ and black energy $p_{{\rm load},i}^{({{\bf b}})}$, such that $p_{{\rm load},i}=p_{{\rm load},i}^{(\textcolor{nicegreen}{{\bf g}})}+p_{{\rm load},i}^{({{\bf b}})}$. The dual pricing dispatch (DPD) clearing routine is given by the following program where, for notational simplicity, the bid and offer set indices (e.g., $\forall j\in\mathcal{B}_{i}$, $\forall l\in\mathcal{O}_{k}$) have been omitted, but are implicitly assumed: 
{\small
\begin{subequations}\label{eq:dpd_si}
\begin{align}\max_{p_{{\rm load}},p_{{\rm gen}}}\quad & \sum_{i\in\mathcal{L}}\left(p_{{\rm load},i}\times c_{{\rm load},i}\right)-\sum_{k\in\mathcal{G}^{(\textcolor{nicegreen}{{\bf g}})}}\left(p_{{\rm gen},k}^{(\textcolor{nicegreen}{{\bf g}})}\times c_{{\rm gen},k}^{(\textcolor{nicegreen}{{\bf g}})}\right)-\sum_{k\in\mathcal{G}^{({\bf b})}}\left(p_{{\rm gen},k}^{({\bf b})}\times c_{{\rm gen},k}^{({\bf b})}\right)+\sum_{i\in\mathcal{L}}\alpha_{i}p_{{\rm load},i}^{(\textcolor{nicegreen}{{\bf g}})}\\
{\rm s.t.}\quad & \forall i\in\mathcal{L},\;\forall k\in\mathcal{G},\;\forall n\in\mathcal{N}:\\
 & \underline{p}_{{\rm gen},k}^{(\textcolor{nicegreen}{{\bf g}})}\le p_{{\rm gen},k}^{(\textcolor{nicegreen}{{\bf g}})}\le\overline{p}_{{\rm gen},k}^{(\textcolor{nicegreen}{{\bf g}})},\quad\forall k\in\mathcal{G}^{(\textcolor{nicegreen}{{\bf g}})}\\
 & \underline{p}_{{\rm gen},k}^{({\bf b})}\le p_{{\rm gen},k}^{({\bf b})}\le\overline{p}_{{\rm gen},k}^{({\bf b})},\quad\forall k\in\mathcal{G}^{({\bf b})}\\
 & \underline{p}_{{\rm load},i}\le p_{{\rm load},i}\le\overline{p}_{{\rm load},i},\quad\forall i\in\mathcal{L}\\
 & \underline{p}_{{\rm load},i}\le p_{{\rm load},i}^{(\textcolor{nicegreen}{{\bf g}})}\le\overline{p}_{{\rm load},i},\quad\forall i\in\mathcal{L}\\
 & \underline{p}_{{\rm load},i}\le p_{{\rm load},i}^{({\bf b})}\le\overline{p}_{{\rm load},i},\quad\forall i\in\mathcal{L}\\
 & p_{{\rm load},i}^{(\textcolor{nicegreen}{{\bf g}})}+p_{{\rm load},i}^{({\bf b})}=p_{{\rm load},i}\label{subeq:energyload},\quad\forall i\in\mathcal{L}\\
 & \sum_{i\in\mathcal{L}_{n}}p_{{\rm load},i}+\sum_{m\in\Lambda_{n}}B_{nm}(\delta_{n}-\delta_{m})-\left(\sum_{k\in\mathcal{G}_{n}^{(\textcolor{nicegreen}{{\bf g}})}}p_{{\rm gen},k}^{(\textcolor{nicegreen}{{\bf g}})}+\sum_{k\in\mathcal{G}_{n}^{({\bf b})}}p_{{\rm gen},k}^{({\bf b})}\right)=0:\;\lambda_{n}\\
 & \underline{p}_{nm}\le B_{nm}(\delta_{n}-\delta_{m})\le\overline{p}_{nm}\\
 & \sum_{i\in\mathcal{L}}p_{{\rm load},i}^{(\textcolor{nicegreen}{{\bf g}})}-\sum_{k\in\mathcal{G}^{(\textcolor{nicegreen}{{\bf g}})}}p_{{\rm gen},k}^{(\textcolor{nicegreen}{{\bf g}})}=0:\;\lambda_{\textcolor{nicegreen}{g}}\label{subeq:lambda_{g}reen}.
\end{align}
\end{subequations}}In this formulation, each bidding load serving entity submits an $\alpha_i \geq 0$, based on how much they value green energy over black across all bid blocks:
\[
\alpha_{i}\rightarrow\text{marginal value of \textcolor{nicegreen}{green} over black (\$/MWh)}.
\]
The objective function is then rewarded, via $\sum_{i\in\mathcal{L}}\alpha_{i}p_{{\rm load},i}^{(\textcolor{nicegreen}{{\bf g}})}$, based on how much green energy is apportioned to each load. Finally, \eqref{subeq:lambda_{g}reen} is a green power balance equation. This equation ensures that all produced green power is properly proportioned to loads (i.e., if only 1 MWh of green is produced, then only 1 MWh of green can be assigned to the loads). In this way, every drop of green generation is allocated to a load.

Since $\alpha_i \geq 0$, green bids will be higher than the bids for black energy and so generally, the maximization of the objective function will push for releasing more green energy. This will raise clearing prices for green energy. In the case where $\alpha_i = 0, \,\forall i$, the dual dispatch framework relaxes to the standard formulation of a market clearing algorithm as given by \eqref{eq:dammarketclearing}. 
As far as the authors know, this is the only model that allows consumers to place two concurrent bids, a green and a black bid, across all bid blocks, while preserving market clearing program's linearity and convexity.

\subsubsection*{Calculating Locational Marginal Prices (LMPs)} In order to compute the LMPs, we formulate the Lagrangian associated with \eqref{eq:dpd_si}:
{\small
\begin{align}
\mathcal{L}=\sum_{n}\lambda_{n}\left(\sum_{i\in\mathcal{L}_{n}}p_{{\rm load},i}+\sum_{m\in\Lambda_{n}}B_{nm}(\delta_{n}-\delta_{m})-\left(\sum_{k\in\mathcal{G}_{n}^{(\textcolor{nicegreen}{{\bf g}})}}p_{{\rm gen},k}^{(\textcolor{nicegreen}{{\bf g}})}+\sum_{k\in\mathcal{G}_{n}^{({\bf b})}}p_{{\rm gen},k}^{({\bf b})}\right)\right)+\lambda_{\textcolor{nicegreen}{g}}\left(\sum_{i\in\mathcal{L}}p_{{\rm load},i}^{(\textcolor{nicegreen}{{\bf g}})}-\sum_{k\in\mathcal{G}^{(\textcolor{nicegreen}{{\bf g}})}}p_{{\rm gen},k}^{(\textcolor{nicegreen}{{\bf g}})}\right)+\cdots
\end{align}}
Next, we hypothesize two sorts of load perturbations at bus $n$: a standard (i.e., non-green) load perturbation $\Delta_{n}$, and a green load perturbation $\Delta_{n}^{(\textcolor{nicegreen}{{\bf g}})}$ (i.e., this load perturbation must be served with green power). Notably, neither of these perturbations are ``paying customers" -- they are simply new infinitesimal loads that must be served. The LMPs may be calculated by taking the gradient of the load perturbations with respect to the individual perturbations.
\begin{align}
{\rm LMP}_{n}^{({\bf b})} & \triangleq\frac{\partial\mathcal{L}}{\partial\mathcal{L}\Delta_{n}}=\lambda_{n},\;\,\qquad\forall n\in\mathcal{N}\\
{\rm LMP}_{n}^{(\textcolor{nicegreen}{{\bf g}})} & \triangleq\frac{\partial\mathcal{L}}{\partial\mathcal{L}\Delta_{n}}=\lambda_{n}+\lambda_{\textcolor{nicegreen}{g}},\;\forall n\in\mathcal{N},
\end{align}
where ${\rm LMP}_{n}^{({\bf b})}$ is the black LMP and ${\rm LMP}_{n}^{(\textcolor{nicegreen}{{\bf g}})}$ the green LMP. $\lambda_{\textcolor{nicegreen}{g}}$ is the dual variable of the green power balance equation, where the total amount of generated green energy should be equal to the total green energy received by the loads. It therefore can be thought of an a ``global'' cost of supplying an extra unit of green energy. Accordingly, all green LMPs are equal to black LMPs, but shifted by the global cost. Interestingly, this global cost will always be positive, regardless of network configuration (see following remark). Thus, green LMPs will always be monotonically higher than black LMPs in our dual dispatch system.

\begin{remark}
    Assuming $\alpha_i\ge 0,\;\forall i$, then $\lambda_{\textcolor{nicegreen}{g}}\ge 0$.
\end{remark}
The nonnegative nature of the green power balance equation shadow price can be shown by transforming the equality constraint of \eqref{subeq:lambda_{g}reen} into an inequality constraint:
$\sum_{i\in\mathcal{L}}p_{{\rm load},i}^{(\textcolor{nicegreen}{{\bf g}})}-\sum_{k\in\mathcal{G}^{(\textcolor{nicegreen}{{\bf g}})}}p_{{\rm gen},k}^{(\textcolor{nicegreen}{{\bf g}})}\le0$. This inequality constraint will always be tight, since assignment of green energy to loads will freely raise the objective function. Since this is an inequality constraint, by the KKT conditions, the associated dual variable will always be nonnegative. Finally, since the inequality constraint is tight, and since its dual variable is nonnegative, this problem's solution is equivalent to the solution to \eqref{eq:dpd_si}, and $\lambda_{\textcolor{nicegreen}{g}}\ge 0$.

















\end{document}